\documentclass[prx,aps,twocolumn,reprint,amsmath,floatfix,amssymb,showpacs,superscriptaddress]{revtex4-1}
\usepackage{graphicx} 
\usepackage[centering,hmargin=19mm,tmargin=27mm,bmargin=27mm]{geometry}
\usepackage{amsmath}
\usepackage{hyperref} 
\usepackage{multirow}
\usepackage{newtxtext}                                        
\usepackage{booktabs}
\usepackage{xcolor} 
\usepackage[normalem]{ulem}
\usepackage[textwidth=1.5cm,textsize=footnotesize]{todonotes}
\hypersetup{colorlinks, 
	linkcolor={blue!75!black!80!yellow},
	citecolor={blue!75!black!80!yellow}, 
	urlcolor={blue!75!black!80!yellow}
	}
\usepackage[cmintegrals]{newtxmath}
\newcommand{\BostonCollege}{Department of Physics, Boston College, Chestnut Hill, MA, USA, 02467}

\makeatletter
\renewcommand\@make@capt@title[2]{%
	\@ifx@empty\float@link{\@firstofone}{\expandafter\href\expandafter{\float@link}}%
	\sffamily{\textbf{#1}}\@caption@fignum@sep#2
}%

\makeatother

\definecolor{darkred}{rgb}{0.7,0.0,0.0}

\definecolor{darkblue}{rgb}{0,0.02,0.45}

\def\bs{\boldsymbol}

\def\vec{\mathbf}
\def\mc{\mathcal}

\newcommand{\be}{\begin{equation}}
\newcommand{\ee}{\end{equation}}
\newcommand{\bea}{\begin{eqnarray}}
\newcommand{\eea}{\end{eqnarray}}
\newcommand{\sbe}{\small\begin{equation}}
\newcommand{\see}{\end{equation}\normalsize}
\newcommand{\sbea}{\small\begin{eqnarray}}
\newcommand{\seea}{\end{eqnarray}\normalsize}
\begin{document} 
\title{{ Signatures of non-Loudon-Fleury Raman scattering in the Kitaev magnet $\beta$-Li$_2$IrO$_3$}}

\author{Yang Yang}
\affiliation{School of Physics and Astronomy, University of Minnesota, Minneapolis, Minnesota 55455, USA}\author{Yiping Wang}\affiliation{\BostonCollege}
\author{Ioannis Rousochatzakis}
\affiliation{Department of Physics, Loughborough University, Loughborough LE11 3TU, United Kingdom}
\author{Alejandro Ruiz}
\affiliation{Department of Physics, University of California, San Diego, California 92093, USA}
\author{James G. Analytis}
\affiliation{Department of Physics, University of California, Berkeley, California 94720, USA}
\affiliation{Materials Sciences Division, Lawrence Berkeley National Laboratory, Berkeley, California 94720, USA}
\author{Kenneth S. Burch}\affiliation{\BostonCollege}
\author{Natalia B. Perkins}
\affiliation{School of Physics and Astronomy, University of Minnesota, Minneapolis, Minnesota 55455, USA}

\date{\today}
\begin{abstract}
We investigate the magnetic excitations of the hyperhoneycomb Kitaev magnet $\beta$-$\text{Li}_2\text{IrO}_3$ by means of inelastic Raman scattering. The spectra exhibits a coexistence of a broad scattering continuum and two sharp low-energy peaks at 2.5\,meV and 3\,meV, with a distinctive polarization dependence. While the continuum is suggestive of fractional quasi-particles emerging from a proximate quantum spin liquid phase, the sharp peaks provide the first experimental signature of the `non-Loudon-
Fleury' one-magnon scattering processes proposed recently [Phys.~Rev.~B~{\bf 104}, 144412 (2021)]. The corresponding microscopic mechanism is similar to the one leading to the symmetric off-diagonal exchange interaction $\Gamma$ (as it involves a combination of both direct and ligand-mediated exchange paths), but is otherwise completely unexpected within the traditional Loudon-Fleury theory of Raman scattering. 
The present experimental verification therefore calls for a drastic reevaluation of Raman scattering in similar systems with strong spin orbit coupling and multiple exchange paths.
\end{abstract}

\maketitle

{\it {Introduction.--}} In recent years, magnetic insulators of 4d and 5d transition metal compounds with bond-directional exchange anisotropies, broadly known as Kitaev materials, have become a rich playground for novel magnetic phases of matter~ \cite{Kitaev2006,Jackeli2009,Jackeli2010,BookCao,Krempa2014,Rau2016,Winter2017,Takagi2019,Trebst2021}. The majority of these systems order magnetically at sufficiently low temperatures~\cite{Takagi2019,Trebst2021}, consistent with theoretical predictions that the Kitaev quantum spin liquid (QSL) phases, that are stabilized by the so-called Kitaev anisotropy $K$, are fragile against weak perturbations~ \cite{Jackeli2010,Schaffer2012,Rau2014,Ioannis2015}. 
However, the usual dominance of the Kitaev coupling renders these materials in relative proximaty to the ideal QSL phases, leading to a general expectation that the magnon modes expected at low energies will coexist with a broad continuum associated with the fractional excitations (spinons) of the nearby QSL phases~\cite{Banerjee2017,Sandilands2015,Glamazda2016,Nasu2016, Rousochatzakis2019,Wulferding2020,Sahasrabudhe2020,Lin2020,Yiping2020,Ruiz2021}.

\begin{figure}[t!]
\includegraphics[width=0.9\columnwidth]{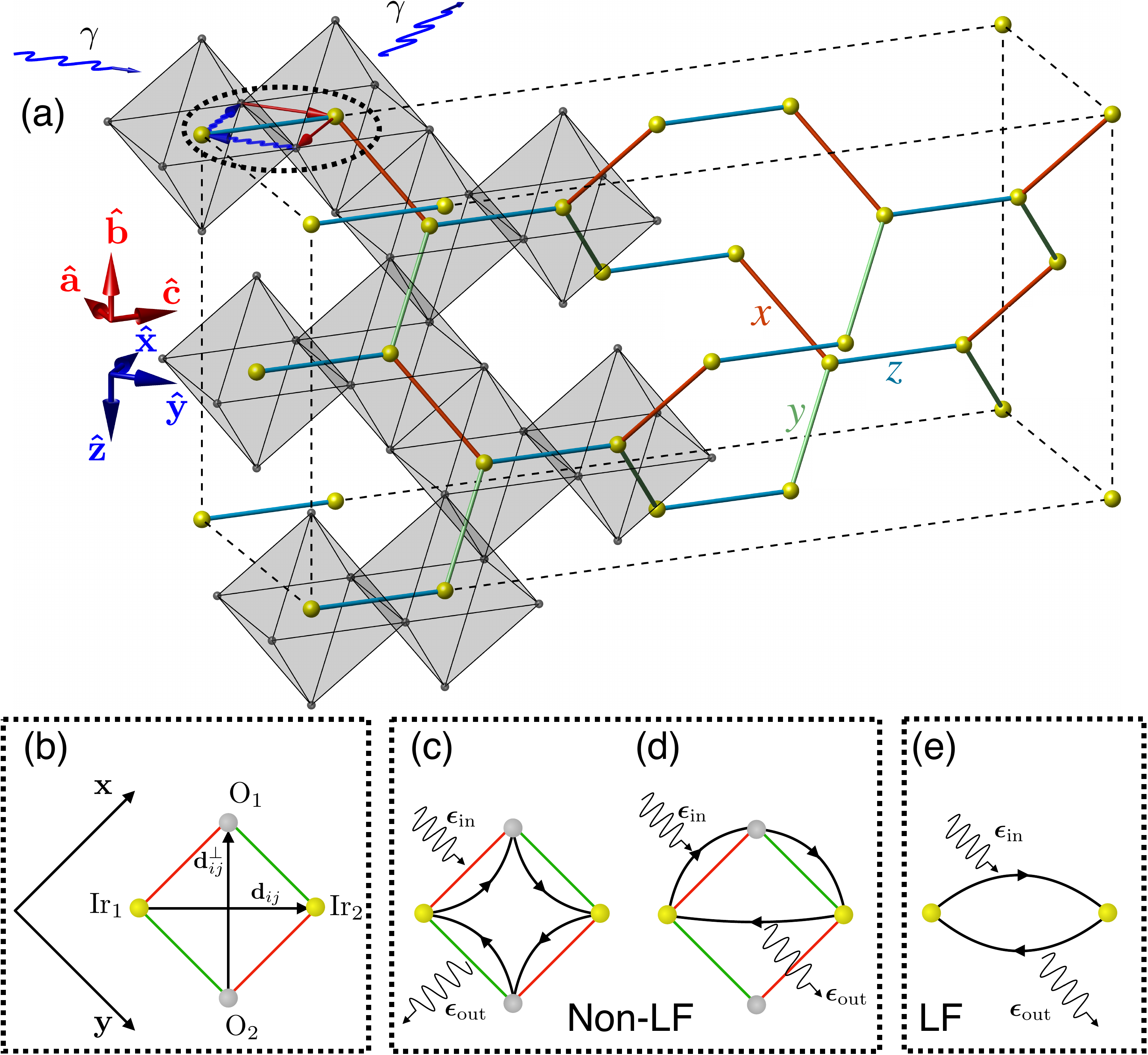}
\caption{(a)  Hyperhoneycomb network of Ir$^{4+}$ ions (yellow spheres) in $\beta$-$\text{Li}_2\text{IrO}_3$. Each octahedron denotes a $\text{IrO}_6$ cage. (b) All microscopic processes leading to the effective Hamiltonian on a given bond are confined to the  $\text{Ir}_2\text{O}_2$ plaquette (also highlighted in (a) by a blue dashed circle).  (c-d) `Non-Loudon-Fleury' Raman processes, in which the virtual, photon-assisted,  electron hopping process does not reduce to the effective exchange times an overall polarization dependence, as in typical LF processes (e).
 }
\label{fig:1}
\end{figure}

Here we explore this picture in the hyperhoneycomb Kitaev material $\beta$-$\text{Li}_2\text{IrO}_3$ with inelastic Raman scattering, which is known to be a sensitive probe to single- and multi-particle excitations over sufficiently wide ranges of temperature and energy~\cite{Glamazda2016,Yiping2020,Wulferding2020,Sahasrabudhe2020}.
The $\beta$-Li$_2$IrO$_3$ compound  features an Fddd orthorombic space group, with a hyperhoneycomb lattice of $Ir^{4+}$ ions, each forming an effective $J_{\rm eff}=1/2$ magnetic moment due to strong spin-orbit coupling (SOC)~\cite{Biffin2014a,Takayama2015,Ruiz2017,Ruiz2020,Majumder2019,Ruiz2021}.
As shown in Fig.~\ref{fig:1}\,(a), the $Ir^{4+}$ ions form zigzag chains (red and green bonds) running alternatively along (${\bf a}$-${\bf b}$, ${\bf a}$-${\bf b}$) and (${\bf a}$+${\bf b}$, ${\bf a}$+${\bf b}$) directions. 
At zero-field and below $T_I\!=\!38$~K, the system shows an incommensurate (IC) order with counter-rotating spin sublattices and propagation wavevector ${\bf Q}\!=\!(0.574,0,0)$ in orthorhombic units~\cite{Biffin2014a,Takayama2015}. This  complex order results from the competition among various bond-dependent anisotropic exchange interactions. Similarly to all other Kitaev materials~\cite{Takagi2019,Trebst2021}, edge-sharing IrO$_6$  octahedra in $\beta$-$\text{Li}_2\text{IrO}_3$ provide 90$^{\circ}$ paths for the dominant bond-directional, Ising-like Kitaev interaction among magnetic moments~\cite{Jackeli2009,Jackeli2010}. %with  the  exchange  easy  axis  which depends  on  the  spatial  orientation  of  the bond 
Besides the dominant Kitaev anisotropy, $\beta$-$\text{Li}_2\text{IrO}_3$  features additional interactions,  such as the nearest neighbor (NN) Heisenberg interaction $J$ and  the symmetric component of the  NN off-diagonal exchange coupling, commonly referred to as the $\Gamma$ interaction~\cite{Rau2014,Katukuri2014,IoannisGamma}.

In agreement with earlier studies by Glamazda {\it et. al.}~ \cite{Glamazda2016}, our experimental results for the Raman susceptibility reveal a broad scattering continuum that survives in a wide temperature range up to 100\,K, well above $T_I$. 
Our analysis of the $T$ dependence of this continuum and the evolution of its spectral weight, as well as a comparison with theoretical calculations, suggests that this  continuum
%, non-phononic background 
is not associated with the magnon excitations of the low-$T$ ordered phase. Rather, the continuum is more consistent with spinons of the proximate Kitaev spin liquid phase, thus reinforcing the magnon-spinon dichotomy picture already advocated previously for this material by a resonant inelastic X-ray scattering  (RIXS) study~\cite{Ruiz2021}.

In addition to the continuum background, however, we have observed  two sharp, low-energy peaks below T$_I$, which were not resolved in Ref.~\cite{Glamazda2016}. These peaks appear in cross polarization and not in the parallel polarization, and furthermore disappear above $T_I$. 
A direct comparison of these findings with the recently~\cite{Yang2021} revised theory of Raman scattering, applicable to Kitav-like Mott insulators with strong SOC, reveals that these peaks are in fact {\it the first experimental signature} of `non-Loudon-Fleury' magnon scattering processes.
More specifically, according to Ref.~\cite{Yang2021}, the leading contributions to the Raman vertex $\mc{R}$ [which enters the Raman intensity $\mc{I}(\Omega)\!\propto\!\int\!d t ~e^{i \Omega t} \langle \mc{R}(t)\mc{R}(0)\rangle$, where $\Omega\!=\!\omega_{\rm in}\!-\!\omega_{\rm out}$ is the total energy transfer %(in units of $\hbar=1$)
and $\langle \cdots\rangle$ denotes thermal averaging], contains significant terms arising from microscopic photon-assisted tunneling processes [Fig.~\ref{fig:1}\,(c-d)] beyond those [Fig.~\ref{fig:1}\,(e)] appearing in the traditional Loudon-Fleury (LF) theory~\cite{LF1968,Shastry1990}.
Among these, the virtual  processes of Fig.\ref{fig:1}\,(d), which involve both direct and ligand mediated paths, are of similar type with the ones leading to the symmetric off-diagonal interaction $\Gamma$, but, in the Raman vertex, they take the form of a bond-directional magnetic dipole term. Such terms are responsible for the appearance of sharp, one-magnon Raman peaks with distinctive polarization dependence, and are otherwise not expected in the traditional LF theory~\cite{Yang2021}.

{\it {Crystal growth, handling and characterization.--}} High-quality single crystals of $\beta$-Li$_2$IrO$_3$ were grown by a vapor transport technique. Ir (99.9\% purity, BASF) and Li$_2$CO$_3$ (99.999 \% purity, Alfa-Aesar) powders were ground and pelletized at 3,000 psi in the molar ratio of 1:1.05. The pellets were placed in an alumina crucible, reacted for 12 h at 1,050\textdegree C, and then cooled down to room temperature at 2 \textdegree C/h to yield single crystals which were then extracted from the reacted powder. $\beta$-Li$_2$IrO$_3$ crystallizes in the orthorhombic Fddd space group and average 105$\times$150$\times$300 $\upmu$m$^3$ in size.

{\it {Raman spectroscopy setup.--}} The Raman spectra presented here were obtained on a custom built, low temperature microscopy setup~\cite{Tian2016RSI,Wang2020KitRam,Gavin_PRX}. A 532 nm excitation laser, whose spot has a diameter of 2 $\mu m$, was used with the power limited to 10 $\mu$W to minimize sample heating while allowing for a strong enough signal. The absence of laser induced heating was crucial to ensure the ordered state is achieved, and is confirmed via stokes/antistokes analysis as well as the appearance of magnons at the appropriate temperature. The single crystal was mounted by silver paint onto a copper sample holder and vacuum transferred onto \emph{xyz} stage in in the cryostat~\cite{Tian2016RSI}. At both room and base temperature (10 K), the reported spectra were averaged from three spectra in the same environment to ensure reproducibility. The spectrometer had a 2400 g/mm grating, with an Andor CCD, providing a resolution of $\approx 1$\,cm$^{-1}$. Dark counts are removed by subtracting data collected with the same integration time with the laser blocked. To minimize the effects of hysteresis from the crystal structural transition, data was taken by first cooling the crystal to base temperature and then heating to the target temperature.

%Furthermore, a recently developed wavelet based approach was employed to remove cosmic rays~\cite{Tian2016dispike}.

\begin{figure}
\includegraphics[width=1\columnwidth]{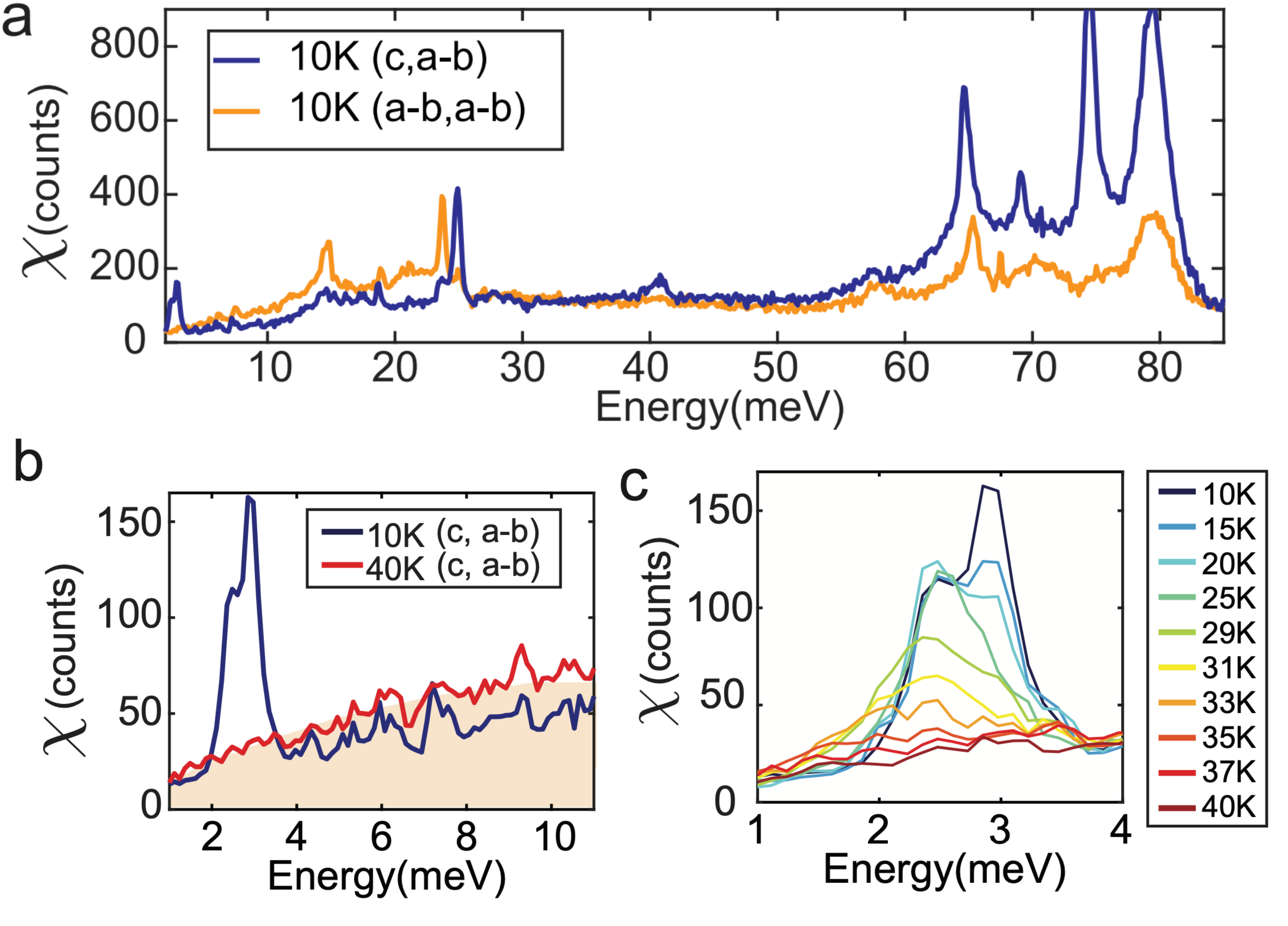}
\caption{
(a) Raman susceptibility of $\beta$-LiIrO$_3$ at 10 K, orange line shows (\textbf{a-b, a-b}) polarization, blue line is (\textbf{c, a-b}) polarization. 
(b) Comparison of Raman susceptibility in the  (\textbf{c,a-b}) channel at 10 K (blue) and 40 K (red).
(c) Temperature dependence of the two low-energy peaks M1 and M2 seen in the (\textbf{c, a-b}) channel}.
\label{fig:2}
\end{figure}

{\it {Results.--}} Figure\,\ref{fig:2}\,(a) shows the 10\,K Raman susceptibility measurement at cross (\textbf{c, a-b}) and parallel (\textbf{a-b, a-b}) polarizations. The notation 
(\textbf{c, a-b}) and (\textbf{a-b, a-b})
refer to the incident and scattered beam polarizations in the orthorhombic reference frame of the crystal structure. 
As the Raman intensity of Stokes and anti-Stokes scattering can be described using $I_{Stokes} = \chi (n_B+1)$ and $I_{anti-Stokes} = \chi (n_B)$, we extracted the Raman susceptibility by dividing the measured intensity by the appropriate Bose function.

{\it {i) Phonon modes.--}} In both polarizations, the spectra show a number of sharp
peaks superimposing a continuum background.
The very sharp peaks appearing above  $\sim$\,25\,meV can be readily identified as optical phonon modes (obeying the selection rules of the \textit{Fddd} space group),  as was  analysed previously  in \cite{Glamazda2016}. 
%
%
%lowest optical phonons begin around 24 meV (Lemmens, Glamazda) next ones start above 40-50?\\
%
%
%
Among them, several peaks have pronounced asymmetric line shapes, which can be ascribed to Fano resonances~\cite{Fano1961} due to the coupling of the optical phonons to the underlying continuum of non-phononic origin. 
A similar asymmetry of the low-energy phonon line shapes has been extensively discussed in studies of bulk $\alpha$-RuCl$_3$, both experimentally~\cite{Sandilands2015,Mai2019,Lin2020,Wulferding2020,Sahasrabudhe2020} and theoretically~ \cite{Metavitsiadis2021,Feng2021}.

%{\cgr In particular, it has been suggested that the temperature evolution of the Fano lineshape of the phonon peaks overlapping with the continuum provides  observable signatures of  fractionalized excitations  characteristic of the underlying spin liquid phase.}{\cbl ** not needed? plus the sentence `underlying spin liquid phase' is incorrect and  confusing. I suggest to remove everything in green**}

%(c) Comparison of  Raman susceptibility  in the  (\textbf{c, a-b}) (shown by blue line) and (\textbf{a-b, a-b}) (shown by orange line) polarization at 10 K. The shade region is a guide to the eye and indicates the continuum contribution.{\color{red} The low energy upturn in (\textbf{a-b, a-b}) is the Rayleigh scattering edge.}

{\it ii) Sharp low-energy peaks.--}
We now turn to low energies and low temperatures, where coherent magnons are most likely to appear. In Fig.~\ref{fig:2}\,(b), we plot the Raman spectra  in (\textbf{c, a-b}) polarization both at 10 K, the lowest temperature in our measurements, and at 40 K, slightly above $T_I$. We observe two nearby but well-resolved peaks at very low energy. Importantly, these peaks appear only in the cross polarization, and are absent in the (\textbf{a-b, a-b}) data of Fig.~\ref{fig:2}\,(a). 
Moreover, at 10 K, the sharp modes and the underlying broad continuum coexist, while at 40 K only the broad continuum survives, suggesting that the former comes from magnons. % and the latter comes from long-lived spinons. 

To better understand the low energy sharp features, we focus on the  (\textbf{c, a-b}) polarization and study their temperature evolution, by performing the Raman scattering with small step temperature increase, see data in Fig.~\ref{fig:2}\,(c). At 10 K, the two peaks are centered around 2.5 meV (M1) and 3 meV (M2), similarly with the two sharp resonances, centered around  2.1 meV and 3 meV, that were previously observed  in the THz spectra~\cite{Majumder2020}. Interestingly the intensity of the M2 peak is larger than the intensity of the M1 peak. However, the two peaks exhibit very different temperature evolution. From 10 K to 29 K, the  intensity of the M2 peak decreases with temperature and is merged into the high energy tail of M1, while M1 increases from 20 K to 25 K, and then starts decreasing and softening until it disappears as we reach $T_I$.

\begin{figure}[!t]
\includegraphics[width=0.9\linewidth]{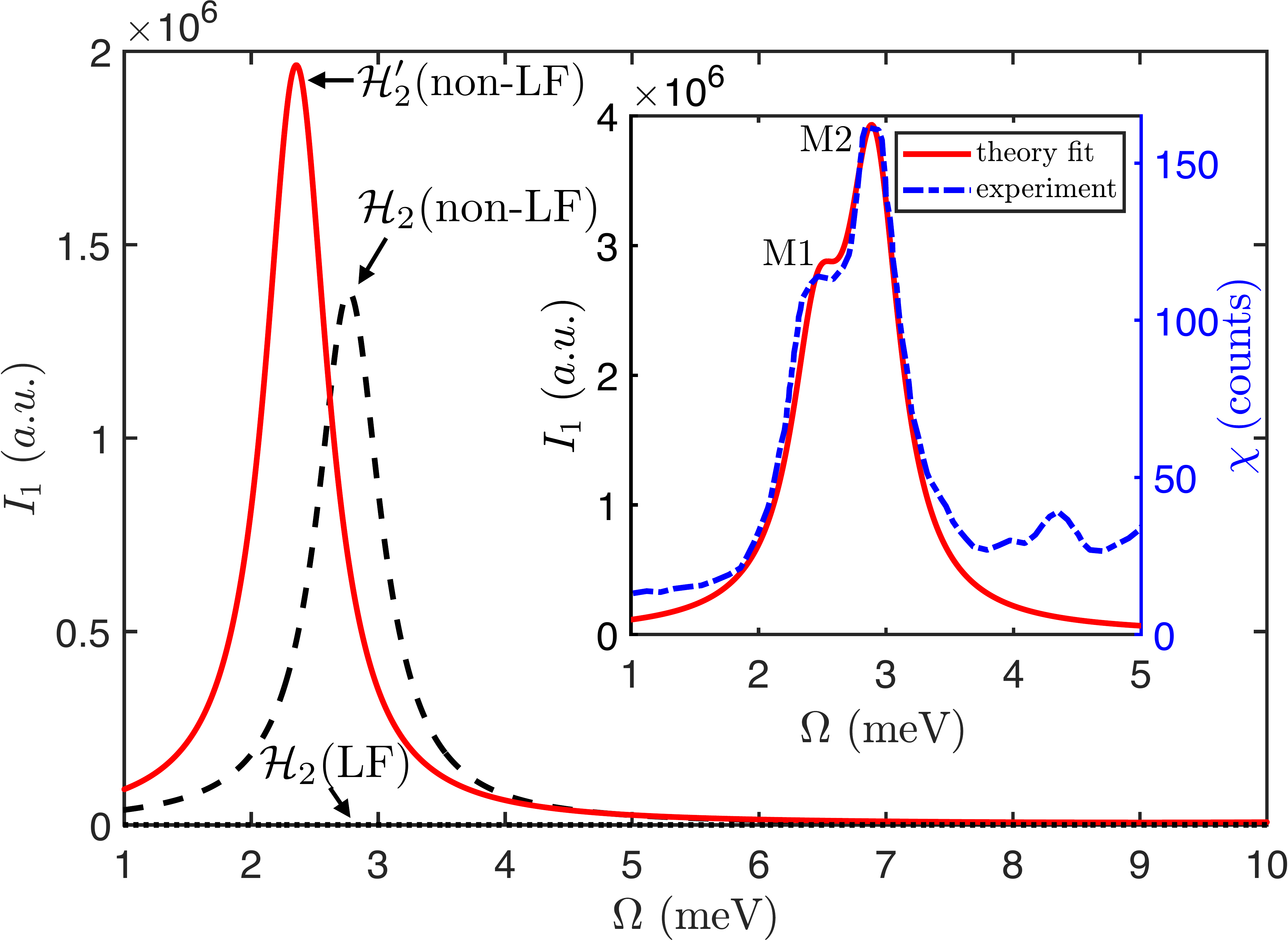}
\caption{
The one-magnon Raman response computed within the non-LF theory, at the level of linear spin wave theory (black dashed line) or with magnons renormalized by the quartic interactions $\mathcal{H}_4$ only (red solid line) [see detailed discussion in the SM], shows one low-energy sharp peak feature in the $(\mathbf{c},\mathbf{a}-\mathbf{b})$ polarization channel.
In contrast, the LSW theory with the LF Raman operator gives no low-energy features (black dotted line,  not visible because the intensity vanishes). 
The inset shows the fit of the low-energy peaks M1 and M2 to the phenomenological model discussed in the SM.}
\label{fig:3a}
\end{figure}

The fact that the two low-energy peaks only exist below $T_I$ implies that they can be assigned to magnons. 
To establish this we employ the recently revised theory of Raman scattering discussed above. As shown in Ref.~\cite{Yang2021}, for the case of $\beta$-Li$_2$IrO$_3$ (and for $\Omega \ll \omega_{\rm in,\, out}$), the non-LF terms of Fig.\ref{fig:1}\, (d) give rise to a sharp, one-magnon peak in the (\textbf{a-c}) channel.
Figure~\ref{fig:3a} shows this peak for the present case of (\textbf{c,a-b}) polarization, as obtained from a semiclassical expansion around the commensurate ${\bf Q}\!=\!(2/3,0,0)$ approximant state of $\beta$-Li$_2$OIr$_3$, and using the minimal $J$-$K$-$\Gamma$ model [ see Supplementary Material (SM)].
At the level of linear spin-wave (LSW) theory (dashed  black line), the position of the peak is centered around  $\omega_2\simeq2.8$~meV, close to the positions of the observed peaks M1 and M2. 
The same calculation for the \textbf{(a-b, a-b)} channel shows no peak at this energy range, consistent with the experimental results.
This agreement on the position of the peak and its polarization dependence gives strong support to the one-magnon origin of one of the two peaks.

What about {\it the second peak?} To address this question we begin by recognizing that the non-interacting magnon spectrum does in fact feature a second low-energy mode at  $\omega_1\simeq0.34$~meV (this is the mode that `unfolds' to the pseudo-Goldstone mode at the ordering wavevector in the dynamical structure factor, see \cite{Yang2021}) but the calculated Raman intensity of this mode at ${\bf Q}=0$ vanishes.
Magnon anharmonicities [treated at the level of a mean-field decoupling of the quartic interactions (and disregarding the magnon decay processes driven by the cubic terms),  see SM] appear to be able to bring the two low energy magnon modes much closer in energy [the renormalized energies are $\omega_1^{\text{ren}}\simeq2.3$\,meV and $\omega_2^{\text{ren}}\simeq3.1$\,meV], as in experiment, however we still see only one mode with nonzero intensity. 
This indicates that the vanishing of the intensity is due to a phase cancellation related to the commensurate character of the considered approximate ground state, in conjunction with the uniform character of the Raman vertex (similarly to the vanishing of the intensity of the dynamical spin structure factor at the zone center in bi-partite N\'eel antiferromagnets~\cite{Headings2010}).
It is then plausible that incorporating the true IC character of the ordered state, or lower symmetry terms that are inevitably present in the spin Hamiltonian, would remove this phase cancellation of the transition matrix element and render the second magnon mode observable as well. 
A simple phenomenological way to incorporate such a coupling by hand into our semiclassical expansion is discussed in the SM, and can readily deliver a good agreement with experiment, see the inset in Fig.~\ref{fig:3a}. Altogether, this suggests that the observed proximity of the two peaks M1 and M2 can well be a manifestation of strong anharmonicities, which is perhaps not surprising given the non-coplanar ordering and the strong anisotropic interactions in this material.

{\it {iii) Multi-peak structure at intermediate energies.--}}
Unlike the sharp low-energy modes M1 and M2, the origin of the multi-peak structure observed at  intermediate energies cannot be readily identified, especially in the region between 10~meV and 50~meV, where we expect a mixture of one- and two-magnon excitations along with overlapping phonon modes that are difficult to disentangle. 
Specifically, the reference calculations obtained at the LSW level in Ref.~\cite{Yang2021} have revealed a superposition of many one- and two-magnon modes due to the complex, multisublattice nature of the ordered state and the large number of resulting magnon branches~\cite{Ducatman2018,Li2019}. Most notably, the results point to a polarization dependent,  multiple-peak structure centered around 12~meV and another around 21~meV, along with a broad (but still structured) two-magnon continuum  between 15 meV and 45 meV. These features can qualitatively account  for some of the structures seen in the experimental data.
However, a more accurate description must take into account the effects of spin-wave anharmonicities and magnon decays, which are expected to play a nontrivial role at this intermediate energy range,  given the non-coplanar ordering and the strong off-diagonal $\Gamma$ couplings, that are additionally the source of finite-state interactions~ \cite{Winter2017nc}.

\begin{figure}[!t]
\centering
\includegraphics[width=1\columnwidth]{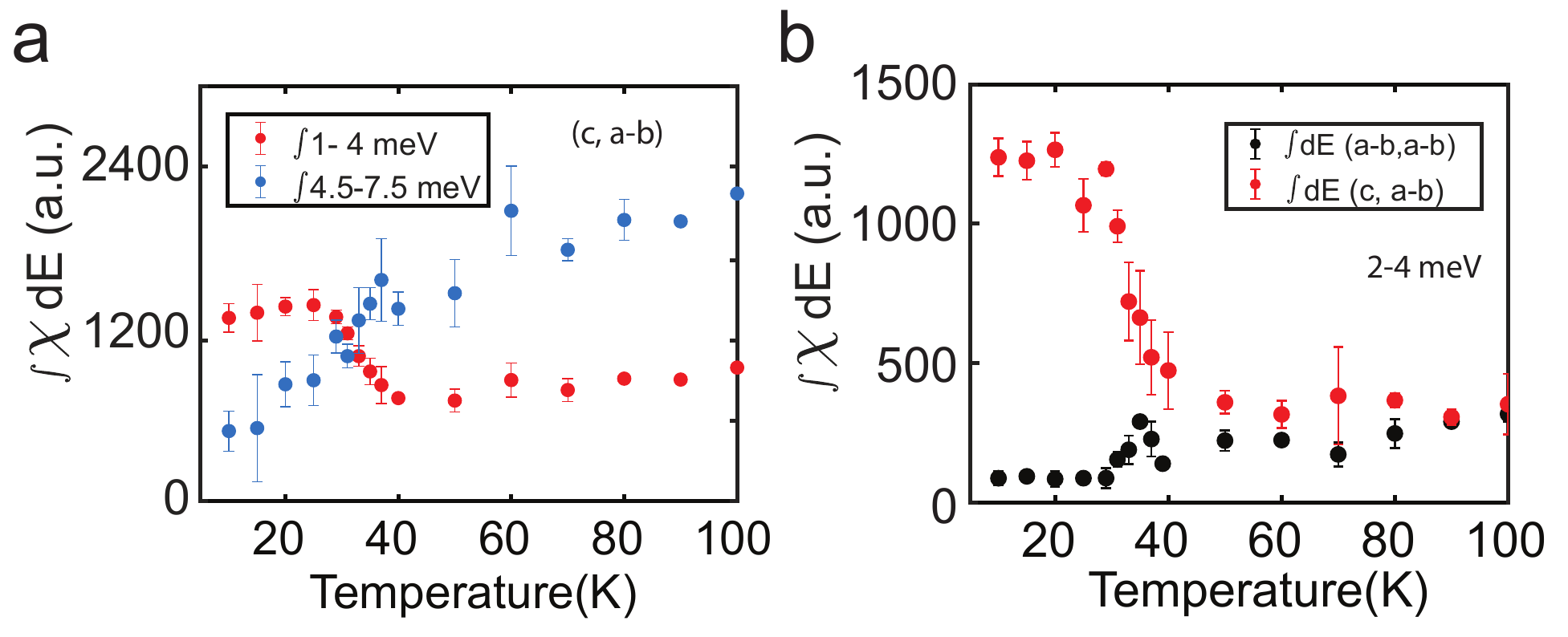}
\caption{ a)  Integrated Raman susceptibility, or spectral weight (SW) in (\textbf{c, a-b}) polarization, as a function  of $T$:
the red dots show the SW from 1 to 4 meV, which includes both M1 and M2 modes and the blue dots represent the SW from 4.5 to 7.5 meV, which  incorporates a 3 meV interval of the broad continuum with no magnon contribution. 
(b) The  SW from 2 to 4 meV  vs $T$ for (\textbf{a-b, a-b})  (black) and (\textbf{c, a-b})  (red)  polarizations. 
}
\label{fig:4}
\end{figure}

{\it {iv) Magnetic continuum.--}}
Let us now return to the continuum background seen in the data. One of the most notable features of this continuum is that it covers a wide energy range, extending all the way down to zero energy, well below the onset of the two-magnon continuum expected for the IC ordered state~\cite{Yang2021}.
Moreover, as mentioned above, 
unlike the sharp low-energy modes M1 and M2, which disappear at $T_I$, the broad continuum persists well above $T_I$, see Figs.~\ref{fig:2}\,(b, c). 
In fact, as we analyse further in Fig.~\ref{fig:4}, the broad continuum persists in a wide temperature range, extending up to $\sim$\,100\,K, well above $T_I$. The presence and very weak $T$-dependence of the continuum as we cross $T_I$ should be contrasted to what happens e.g., in unfrustrated magnets, where, with increasing $T$, the spectral weight broadens and shifts to lower energies and finally evolves to quasielastic scattering from overdamped short-range magnetic fluctuations above the ordering temperature~\cite{Martin1977,Liu1993,Gretarsson2016,Glamazda2016}.
This suggests that in $\beta$-Li$_2$IrO$_3$ much of the continuum background, and especially the one persisting down to zero energy  (where neither magnons nor phonons are expected as mentioned above), is not related to magnons. 
On the other hand, such a continuum Raman response would be generally expected for the fractional quasiparticles of the proximate Kitaev model~\cite{Knolle2014,Perreault2015,Nasu2016,Yamaji2018,Rousochatzakis2019,Gabor2019,Glamazda2016,Sandilands2015,Banerjee2017}.

To explore this further, we follow previous studies and proceed to analyse the integrated Raman susceptibility, or the spectral weight (SW). Figure~\ref{fig:4} (a) 
shows the $T$-dependence of the SW in the  (\textbf{c, a-b}) polarization  in two energy ranges: one between 1-4~meV, which includes both M1 and M2  modes and the underlying continuum, and the other between 4.5-7.5\,meV from the continuum only. We can see that the  lower-energy SW, governed primarily by the two low-energy magnon modes, rapidly decreases with $T$ until it reaches  $T_I$, above which it shows nearly no $T$-dependence. By contrast, the higher-energy SW, which comes solely from the continuum background, keeps increasing with $T$ even above $T_I$, until it  roughly levels off around 100 K. 
Next, we compare the low-energy SW obtained in (\textbf{c, a-b})  and (\textbf{a-b, a-b}) polarizations, see Fig.~\ref{fig:4}\,(b).  At low  $T$, the (\textbf{c, a-b}) SW is significantly larger than the (\textbf{a-b, a-b}) SW, due to the presence of the low-energy magnon modes in the former channel. 
Above $T_I$, the two SWs,  both originating solely from the continuum background, saturate to some temperature independent values, with the (\textbf{c, a-b}) SW being slightly larger than the (\textbf{a-b, a-b}) SW, consistent with the theory prediction for the pure Kitaev model on the hyperhoneycomb lattice~ \cite{Perreault2015}. 
Taken together,
the results from the SW analysis points to a systematic SW transfer from magnons to the continuum background as we approach $T_I$. Conceptually, this ties in with the intuitive picture of magnons turning into pairs of deconfined spinons of the proximate Kitaev phase as we enter the paramagnetic phase.

%This can also be seen by the shading in Fig.~\ref{fig:2}\,(b), which indicates that the broad continuum at 40\,K carries overall larger spectral weight compared to the one at 10\,K.
%

{\it {Summary.--}} 
Our inelastic Raman scattering 
data provides significant insights for the magnetic response of the hyperhoneycomb Kitaev magnet $\beta$-Li$_2$IrO$_3$ in a wide energy and temperature range. In our study, we have provided evidence that while the observed continuum background is likely of magnetic origin, it {\it cannot} be associated with the magnon excitations of the low-temperature ordered phase. The systematic transfer of spectral weight from the sharp low-energy peaks M1 and M2 (which we have established to be magnons) to the continuum background as we heat up the system is consistent with the interpretation of the continuum in terms of fractional excitations emerging from the proximate spin liquid phase, as discussed in previous studies~\cite{Sandilands2015,Glamazda2016,Banerjee2017,Ruiz2021,Yamaji2018,Rousochatzakis2019}.

Turning to the sharp low-energy peaks M1 and M2, we have  demonstrated numerically that their temperature and polarization dependence can only be explained by extending the traditional Loudon-Fleury theory of Raman scattering, where the contribution $\mc{R}_{ij}$ to the Raman vertex from a given bond $\mathbf{d}_{ij}$ is given by the corresponding superexchange Hamiltonian $\mc{H}_{ij}$  weighted by a bond-specific polarization-dependent factor. 
In reality, $\mc{R}_{ij}$ can have a different functional form than $\mc{H}_{ij}$, as the various electron hopping paths each come with their own, non-equivalent polarization factors~\cite{Yang2021}.
In the present case, we have shown that the observed peaks M1 and M2 verify the existence of the `non-Loudon-Fleury' dipole terms that arise from the interplay of direct and ligand-mediated hopping, similarly to the exchange terms leading to the $\Gamma$ interaction. 
This experimental verification therefore marks a drastic change of paradigm for the understanding of Raman scattering in materials with strong SOC and multiple exchange paths.

{\it Acknowledgments:} We thank Mengqun Li for for earlier collaborations on related topics.
Y.Y. and N.B.P. acknowledge the support from the U.S. Department of Energy, Office of Science, Basic Energy Sciences under Award No. DE-SC0018056. The work done by Y.W. was supported by the Office of Naval Research under Award number N00014-20-1-2308. K.S.B. is grateful for the support of the US Department of Energy(DOE), Office of Science, Office of Basic Energy Sciences under award no. DE-SC0018675.
The work of I.R. was supported by the Engineering and Physical Sciences Research Council [grant number EP/V038281/1].
Work by JGA and AR was supported by the Department
of Energy Early Career Program, Office of Basic Energy Sciences, Materials Sciences and
Engineering Division, under Contract No. DE-AC02-05CH11231 for crystal growth.
A.R. also acknowledges support from the University of California President Postdoctoral Fellowship Program.

\bibliography{Reference}
%\bibliography{bib} 
% \end{document}
%%%%%%%%%%%%%%%%%%%%%%%%%%%%%%%%%%%%%%%%%%%%%%%%%%%%%%%%%
%%%%%%%%%%%%%%%%%%%%%%%%%%%%%%%%%%%%%%%%%%%%%%%%%%%%%%%%%
\clearpage
\onecolumngrid

\pagebreak
\appendix
\setcounter{figure}{0}
\setcounter{equation}{0}
\renewcommand{\theequation}{\arabic{equation}}

\section*{Supplemental Material}

%It has been suggested  that the magnetism of  $\beta$-$\text{Li}_2\text{IrO}_3$ can be accurately described by the minimal  $J$-$K$-$\Gamma$ model \cite{Lee2015,Lee2016,Ducatman2018,Rousochatzakis2018,Li2019,Li2020}.  Moreover, it has been shown that the IC order can be understood as a long-distance twisting of a nearby commensurate state with ${\bf Q}\!=\!(2/3,0,0)$ (in units of $2\pi/a$). This state is amenable to a semi-analytical treatment of the problem \cite{Ducatman2018,Rousochatzakis2018,Li2019,Li2020}, with results that are consistent with almost all experimental findings collected so far, both in zero and at finite fields ~\cite{Biffin2014a,Ruiz2017,Majumder2019,Majumder2020,Ruiz2020,Ruiz2021}.

In  this supplemental material, we provide technical details of  spin-wave theory for the minimal Hamiltonian describing the properties of $\beta$-Li$_2$IrO$_3$ (Section S1),  the non-Loudon-Fleury formalism of magnetic Raman scattering (Section S2), and  one-magnon Raman response in $\beta$-Li$_2$IrO$_3$ (Section S3).

\subsection*{S1. Spin-wave theory for the minimal Hamiltonian describing the properties of $\beta$-Li$_2$IrO$_3$}\label{app:SWT}

In order to calculate the Raman response from $\beta$-Li$_2$IrO$_3$, 
we need first to compute the spectrum of magnetic excitations in  $\beta$-Li$_2$IrO$_3$. However, it is not a trivial task since the magnetic ground state of $\beta$-Li$_2$IrO$_3$ is a counter-rotating spiral characterized by an incommensurate wave vector. As such the semiclassical expansion around such state contains Umklapp magnon scattering processes, which leads to an intractable, spin-wave Hamiltonian matrix of infinite size. To circumvent this obstacle we exploit the idea that such inhomogeneous states typically represent a long-wavelength twisting of a nearby commensurate state. We also assume that the correlations at short distances of the incommensurate order resemble the ones of this nearby commensurate state. Thus we approximate the excitation spectrum of $\beta$-Li$_2$IrO$_3$  by the excitation spectrum of the simplest nearby commensurate state with counter-rotating moments, the same irreducible representation and similar periodicity with the one observed experimentally. This is obtained by minimizing the classical energy of  the nearest neighbor (NN) $J$-$K$-$\Gamma$ model
  \begin{equation}\label{eq:Hamiltonian}
\mc{H}\!=\!\sum_{\langle ij\rangle_\nu}\left(J\vec{S}_i\cdot\vec{S}_j+K S_i^{\alpha_\nu}S_j^{\alpha_\nu}+\sigma_\nu\,\Gamma(S_i^{\beta_\nu}S_j^{\gamma_\nu}+S_i^{\gamma_\nu}S_j^{\beta_\nu})\right)\,,
\end{equation}
  where ${\bf S}_i$ denotes the pseudo-spin $j_{\text{eff}}\!=\!1/2$ operator at site $i$, $\nu\in\{x,y,z\}$ labels the three different types of NN Ir-Ir  bonds, and  the prefactor $\sigma_{\nu}$ is $+1$ for $x$ and $y'$   and  $-1$ for $x'$ and $y$ bonds (e.g., see Fig.~1 in \cite{Ducatman2018}).  
This state is  characterized by the ${\bf Q}\!=\!\frac{2}{3}\hat{\bf a}$ and
  features six spin sublattices, three along the $xy$-chains and three  along the $x'y'$-chains~\cite{Ducatman2018}.  In order to accommodate  this ${\bf Q}\!=\!\frac{2}{3}\hat{\bf a}$ magnetic order,
    we make use of an enlarged magnetic unit cell  composed of three orthorhombic unit cells along the ${\bf a}$-axis, and thus contains $\mc{N}_{\text{m}}=48$ magnetic sites. 
Subsequently, we express the local spins in terms of the Holstein-Primakoff  bosons  $a_{{\bf R},\mu}^{\dagger}$ and $a_{{\bf R},\mu}$  and expand the Hamiltonian in powers of  $1/S$ about the classical limit. 
 In the linear spin wave (LSW) theory, we only keep the terms that are quadratic in the bosonic operators.  Then going into momentum space, with $a_{{\bf q},\mu}\!=\!\frac{1}{\sqrt{\mc{N}_{\text{m}}}}\sum_{\bf R} e^{i {\bf q}\cdot{\bf R}} a_{{\bf R},\mu}$ (with ${\bf q}$ belonging to the first magnetic Brillouin zone) gives the following quadratic Hamiltonian:
\sbe\label{H2LSW}
\mc{H}_2=%E_{cl}/S+
\sum_{\vec{q}} {\bf x}_\vec{q}^\dagger \cdot {\bf H}_{\vec{q}}\cdot {\bf x}_\vec{q}\; ,
\see
where %$E_{cl}$ is the classical energy, 
$
{\bf x}_\vec{q}=\left( a_{\vec{q},1}\, ,...\, ,a_{\vec{q},\mc{N}_{\text{m}}}\, ,a_{-\vec{q},1}^\dagger\, ,...\, ,a_{-\vec{q},\mc{N}_{\text{m}}}^\dagger \right)^{\text{T}}\,,
$ 
where ${\bf H}_{\vec{q}}$ is a $(2\mc{N}_{\text{m}})\times(2\mc{N}_{\text{m}})$ matrix. The diagonalization of ${\bf H}_\vec{q}$ involves introducing a set of Bogoliubov quasiparticle operators
$
{\bf y}_\vec{q}\!=\!\left( b_{\vec{q},1}\, ,...\, ,b_{\mc{N}_{\vec{q},\text{m}}}\, ,b_{-\vec{q},1}^\dagger\, ,...\, ,b_{-\vec{q},\mc{N}_{\text{m}}}^\dagger \right)^{\text{T}}\,, 
$
obtained from ${\bf x}_\vec{q} $ by a canonical transformation ${\bf x}_\vec{q}=\vec{T_q}\cdot {\bf y}_\vec{q}$, where $\vec{y_q}$  satisfies the bosonic commutation relations $[b_{\vec{q},\mu},b_{\vec{q},\nu}^\dagger]=\delta_{\mu\nu}$, $[b_{\vec{q},\mu},b_{\vec{q},\nu}]=[b_{\vec{q},\mu}^\dagger,b_{\vec{q},\nu}^\dagger]=0$.
This leads to
\be\label{H2Diag}
\mc{H}_{\rm{LSW}}=\mc{H}_{2}  = \sum_{\bf q}\sum_{\mu=1}^{\mc{N}_m} \omega_{\mu,{\bf q}} \Big(
b_{\mu,{\bf q}}^\dagger b_{\mu,{\bf q}} + \frac{1}{2}
\Big)\,,
\ee
where the new bosons $b^\dagger_{\mu,{\bf q}}$ describe the magnon excitations with frequencies $\omega_{\mu,\bf q}$.  Computation with $J\!=\!0.4~\text{meV}$, 
$K\!=\!-18~\text{meV} $ and $\Gamma\!=\!-10~\text{meV}$ gives the lowest energy states $\omega_{1,{\bf q}=0}=0.3443$ meV and $\omega_{2,{\bf q}=0}=2.7687$ meV \cite{Yang2021}, which are different from the energies of the observed M1 and M2 modes. 

 We next consider the corrections to the LSW  theory due to the interactions between the spin waves, which are expected to be non-negligible due to the non-collinearity and complexity  of the period-3 ground state and the presence of bond-anisotropic $\Gamma$-terms. It turned out that an explicit calculation of the magnon-magnon interactions in the   magnetic order with 48 sublattices is a challenging problem.  Thus, we have estimated the effects of magnon-magnon interaction in a simplified manner and   perform only the first-order corrections due to the quartic anharmonicities, since even the computation of the second-order correction due to the cubic anharmonicities  is more involved.  The $\mc{H}_4$  term is obtained by  keeping  higher order terms in the $1/S$ Holstein-Primakoff  expansion of  the Hamiltonian  (\ref{eq:Hamiltonian}). In the magnon picture, the   $\mc{H}_4$ term can be written as
 \begin{align}
     &\mathcal{H}_4=\sum_{\mathbf{R},\langle \mu\mu'\rangle_\nu}\Bigl(L_{1}^{\mu\mu'}a_\mu a_{\mu'}^\dagger a_{\mu'} a_{\mu'}+L_{2}^{\mu\mu'}a_\mu^\dagger a_{\mu'}^\dagger a_{\mu'} a_{\mu'}+\nonumber\\&L_{3}^{\mu\mu'}a_{\mu'} a_{\mu}^\dagger a_{\mu} a_{\mu}+L_{4}^{\mu\mu'}a_{\mu'}^\dagger a_{\mu}^\dagger a_{\mu} a_{\mu}
  %   \nonumber\\+&
    + L_{5}^{\mu\mu'}a_\mu^\dagger a_\mu a_{\mu'}^\dagger a_{\mu'}+\text{h.c.}\Bigl),
 \end{align}
%   \begin{eqnarray}
%     \mathcal{H}_4 &=& \sum_{\langle ij \rangle_\nu}
%     \Bigl[ V^{1\mu\mu'}_{ij,\nu} a_{i,\mu}^{\dagger} a_{i,\mu}a_{j,\mu'}^{\dagger}
%     a_{j,\mu'}  +
%      V^{2\mu\mu'}_{ij,\nu} a_{i,\mu} a_{j,\mu'}^{\dagger} a_{j,\mu'}a_{j,\mu'} + \\\nonumber&&
%   V^{3\mu\mu'}_{ij,\nu} a_{i,\mu}^{\dagger} a_{i,\mu} a_{i,\mu} a_{j,\mu'} + V^{4\mu\mu'}_{ij,\nu}a_{i,\mu}^{\dagger} a_{j,\mu'}^{\dagger}
%     a_{j,\mu'}^{\dagger} a_{j,\mu'} +   \\\nonumber&&V^{5\mu\mu'}_{ij,\nu}a_{i,\mu}^{\dagger} a_{i,\mu}^{\dagger} a_{i,\mu}
%     a_{j,\mu'}^{\dagger} + V^{6\mu\mu'}_{ij,\nu}a_{i,\mu} a_{j,\mu'}^{\dagger} a_{j,\mu'}^{\dagger} a_{j,\mu'}+\\\nonumber&&
%   V^{7\mu\mu'}_{ij,\nu}  a_{i,\mu}^{\dagger} a_{i,\mu} a_{i,\mu} a_{j,\mu'}^{\dagger} + V^{8\mu\mu'}_{ij,\nu}a_{i,\mu}^{\dagger}
%     a_{j,\mu'}^{\dagger} a_{j,\mu'} a_{j,\mu'}+\\\nonumber&&V^{9\mu\mu'}_{ij,\nu} a_{i,\mu}^{\dagger} a_{i,\mu}^{\dagger} a_{i,\mu}
%     a_{j,\mu'}
%     \Bigr],
% \end{eqnarray}
where  $L_{i}^{\mu\mu'}$   are possible four-magnon vertices  on the bond $\langle \mu\mu' \rangle_\nu$ between  spins on sublattices $\mu$ and $\mu'$ from each type of the quartic terms.
 The corrections due to quartic terms are  obtained by decoupling   $\mc{H}_4$ allowing all possible pair averaging. For example, the term associated with $L_{1}^{\mu\mu'}$ can be decoupled by
 \begin{align}
     a_\mu a_{\mu'}^\dagger a_{\mu'} a_{\mu'}\approx&\langle a_\mu a_{\mu'}^\dagger\rangle a_{\mu'} a_{\mu'}+2\langle a_\mu a_{\mu'}\rangle a_{\mu'}^\dagger a_{\mu'}\nonumber\\
     +&2\langle a_{\mu'}^\dagger a_{\mu'}\rangle a_\mu  a_{\mu'}+\langle a_{\mu'} a_{\mu'}\rangle a_\mu a_{\mu'}^\dagger,
 \end{align}
 where the expectation value $\langle \dots\rangle$ is taken with respect to the ground state of $\mathcal{H}_2$, the vacuum of the Bogoliubov quasiparticles. With such decoupling, we obtain additional terms contributing into the quadratic Hamiltonian from $\mathcal{H}_4$, and along with $\mathcal{H}_2$, we have a modified quadratic Hamiltonian $\mathcal{H}_2'$. The diagonalization of $\mathcal{H}_2'$ leads to a new set of magnon modes  with renormalized energies.
 %are renormalized by the interaction through the quartic Hamiltonian $\mathcal{H}_4$. 
 Surprisingly, the renormalized values of the two low-energy modes give us $\omega'_{1,{\bf q}=0}=2.3$ meV and $\omega'_{2,{\bf q}=0}=3.1$ meV, 
   which are in a good agreement with experimental values. This indicates that in order to better understand the low-energy Raman response, the anharmonic terms beyond the LSW theory need to be taken into consideration.

 \subsection*{S2. Non-Loudon-Fleury formalism for Raman scattering}\label{sec:onemagnon}
 
 %\subsubsection{ Derivation of the one-magnon Raman intensity}
 
 Here we briefly review the main points of a recent study revisiting the theory of magnetic Raman scattering in the  spin-orbit coupled Mott insulators~\cite{Yang2021}.
  In particular, it was explicitly demonstrated that in the Kitaev materials\cite{Jackeli2009,Jackeli2010,Sizyuk2014, Rau2014,Winter2017} the Raman operator has contributions  not accounted in the Loudon-Fleury (LF) theory \cite{LF1968}, in which
the contribution $\mc{R}_{ij}$ to the total Raman operator from a bond $\mathbf{d}_{ij}$ is simply given by the super-exchange interactions $\mc{H}_{\text{eff},ij}$ on that bond, weighted by a bond-specific polarization-dependent factor, i.e.,
\begin{align}\label{eqn:LF}
    \mc{R}_{ij}\sim-(\boldsymbol{\epsilon}_{\text{in}}\cdot\mathbf{d}_{ij})(\boldsymbol{\epsilon}_{\text{out}}\cdot\mathbf{d}_{ij})\mc{H}_{\text{eff},ij},
\end{align}
where $\boldsymbol{\epsilon}_{\text{in}}$ and $\boldsymbol{\epsilon}_{\text{out}}$ denote the polarization direction of the incoming and the outgoing light, respectively. The LF process implies the light only couples to the effective path connecting two magnetic ions, which overlooks the explicit electron(hole) superexchange paths. In the systems with multiple non-equivalent super-exchange paths,
the polarization factors, coming from the coupling to light, couple to the first and the last hopping steps, give unequal weights to different paths. Hence, the summation over these paths leads to a Raman operator $\mc{R}_{ij}$ that is, in general, not proportional to $\mc{H}_{\text{eff},ij}$. obtained by summing up the contributions from all possible paths with the equal polarization weight. This leads to the appearance of the contributions in the Raman response that cannot be obtained within the traditional  LF theory~\cite{LF1968,Shastry1990}.

%   \begin{figure}[!t]
% \centering
% \includegraphics[width=\linewidth]{Fig/NonLF_process-crop.pdf}
% \caption{
% (a) The square plaquette that is relevant for deriving the full Raman operator from the superexchange processes between two magnetic ions sharing a `$z$-bond' in $\beta$-Li${}_2$IrO${}_3$. Here we show (b) the direct hopping process. To be distinguished from the Loudon-Fleury process, only the direct hopping between orbitals on two magnetic ions is considered, and the effective hopping has to be considered explicitly as (c) the mediated hopping and (d) the mixed hopping. In latter two paths, the light can also couples to the bond connecting between the magnetic ion(Ir) and the ligand ion(O).\label{fig:S1}}	
% \end{figure}

 The detail derivation of the magnetic Raman response in  the Kitaev materials can be found in Ref.[\onlinecite{Yang2021}].  Based on this derivation, the total Raman operator can be written as %here we  outline  the microscopic processes leading to the Raman response in $\beta$-Li$_2$IrO$_3$. The virtual hopping processes leading to the $J$-$K$-$\Gamma$ model (and the ones contributing to the Raman operator) are confined to a  plaquette consisting of two magnetic ions and two ligand ions shown in  Fig. \ref{fig:Plaquette}.
%There are three different types of paths on such a plaquette: (a) direct hopping between Ir ions  Ir$_1\rightarrow$Ir$_2\rightarrow$Ir$_1$ contributing already at the second order of the perturbation, (b) oxygen-mediated hopping  of the type Ir$_1\rightarrow$O${_{1(2)}}\rightarrow$Ir$_2\rightarrow$O${_{1(2)}}\rightarrow$Ir$_1$ contributing  at fourth order, and (c) mixed direct/oxygen-mediated hopping processes  of the type of the type Ir$_1\rightarrow$O${_{1(2)}}\rightarrow$Ir$_2\rightarrow$Ir$_1$ arising at third order.  The total Raman operator  is the sum over the contributions from all the microscopic processes, i.e.
 \begin{align}
 \label{Raman1}
\mathcal{R} 
= &\sum_{<ij>}\left( \mathcal{R}^{\text{dir}}_{ij} +\mathcal{R}^{\text{med}}_{ij}+\mathcal{R}^{\text{mix}}_{ij}\right),
 \end{align}
which shows that the total Raman response comes from three distinct  microscopic hopping processes: the direct hopping, the oxygen-mediated hopping, and the mixed hopping, just like the microscopic origin of the superexchange Hamiltonian.   For the direct hopping process, there is only one path that directly connects two magnetic ions, so $\mathcal{R}^{\text{dir}}_{ij}$ keeps the LF form of the Raman operator similar to (\ref{eqn:LF}).  For example, on the $z$ bond, we can write
\begin{align}
    \mc{R}^{\text{dir}}_{\langle ij\rangle_z} &= -\zeta(\boldsymbol{\epsilon}_{\text{in}}\cdot\mathbf{d}_{ij})(\boldsymbol{\epsilon}_{\text{out}}\cdot\mathbf{d}_{ij})(J^{(2)}\mathbf{S}_i\cdot\mathbf{S}_j+K^{(2)} S_i^z S_j^z),
\end{align}
where the constant $\zeta=e^2/(\hbar^2 c^2)g_{\text{in}}g_{\text{out}}$ only depends on the incoming and outgoing light frequency.  However, for the oxygen-mediated hopping and the mixed hopping processes, since more than one path is involved, the corresponding  Raman operators will contain terms beyond the LF formalism. For example, on the $z$ bond,  the Raman operators associated with the oxygen-mediated and the mixed hopping processes are given by
\begin{align}\label{eqn:Rmed}
    \mc{R}^{\text{med}}_{\langle ij\rangle_z} &=-\frac{P_{dd}}{4} K^{(4)} S_i^zS_j^z-\frac{P_{d^\perp d^\perp}}{4} \left( J'^{(4)} {\bf S}_i\cdot{\bf S}_j\!+\!K'^{(4)} S_i^z S_j^z \right),\\\label{eqn:Rmix}
    \mc{R}^{\text{mix}}_{\langle ij\rangle_z} &=  -\frac{P_{dd}}{2}\Gamma^{(3)} (S_i^xS_j^y \!+\! S_i^yS_j^x) -\frac{P_{d^\perp d}\!-\!P_{d d^\perp}}{2}i\,h_\Gamma^{(3)}(S_i^z \!+\! S_j^z),
\end{align}
where $K^{(4)}$, $K'^{(4)}$, $J'^{(4)}$, $\Gamma^{(3)}$ and $h_{\Gamma}^{(3)}$ are frequency dependent Raman coupling constants (to be distinguished fro the super-exchange couplings) \cite{Yang2021}, and we also define the polarization factors 
\begin{align}
    P_{dd}\,&\equiv~\zeta(\bs{\varepsilon}_{\text{in}}\cdot\mathbf{d}_{ij})~(\bs{\varepsilon}_{\text{out}}\cdot\mathbf{d}_{ij}),\nonumber\\\nonumber
    P_{d^\perp d^\perp}&\equiv~\zeta(\bs{\varepsilon}_{\text{in}}\cdot\mathbf{d}_{ij}^\perp)~(\bs{\varepsilon}_{\text{out}}\cdot\mathbf{d}_{ij}^\perp),\\
    P_{d d^\perp}\,&\equiv~\zeta
    (\bs{\varepsilon}_{\text{in}}\cdot\mathbf{d}_{ij})~(\bs{\varepsilon}_{\text{out}}\cdot\mathbf{d}_{ij}^\perp),\\
    P_{d^\perp d}\,&\equiv~\zeta(\bs{\varepsilon}_{\text{in}}\cdot\mathbf{d}_{ij}^\perp)~(\bs{\varepsilon}_{\text{out}}\cdot\mathbf{d}_{ij}).\nonumber
\end{align}
%In the limit of the incoming light frequency $\omega_{\text{in}}\rightarrow 0$
%Neglecting the incoming frequency, $J^{(2)}$, $K^{(2)}+K^{(4)}$, $\Gamma^{(3)}$ are the same as the superexchanage coupling constants $J$, $K$, $\Gamma$. Thus, 
We can recognize the first terms in (\ref{eqn:Rmed}) and %https://www.overleaf.com/project/6193d9d85c0a1ac0a32a6d4d
(\ref{eqn:Rmix}) as contributing to the LF form of the Raman operator, as explicitly discussed in \cite{Yang2021}, and the remaining terms corresponding to the non-LF contributions. Among the latter, the term associated with $h_{\Gamma}^{(3)}$ in (\ref{eqn:Rmix}) plays a crucial role in identifying the low-energy one-magnon excitation modes. As it is pointed out in the Fig.3 of the main text, computing the Raman response for $\beta$-Li$_2$IrO$_3$ with only the terms from the LF formalism shows no low-energy one-magnon excitation.

\subsection*{S3. One-magnon Raman response in $\beta$-Li$_2$IrO$_3$}
Knowing the Raman operator, we can compute the Raman intensity as
      \begin{align}
\label{Intensity}
 I(\Omega)\!=\!\frac{1}{\pi}\text{Im}\int\!d t  ~e^{i \Omega t} [-i\langle \mathcal{R}(t)\mathcal{R}(0)\rangle],
 \end{align} 
 where $\Omega=\omega_{\rm in}-\omega_{\rm out}$ is the total energy transferred to the system ($\hbar=1$), and in the following
we assume that $\Omega \ll \omega_{\rm in,\, out}$; $\langle \cdots\rangle$ is the average with respect to the ground state.
As we  discuss above,  we assume  that M1 and M2 modes observed in experiment correspond to the one-magnon  Raman scattering.
While in practice it is difficult to separate the contributions from one-, two- or multiple magnon scatterings, in theory it is straightforward. To this end, we
expand the Raman operator $\mathcal{R}$ in powers of $1/\sqrt{S}$, 
\begin{align}\label{eq:ExpandR}
\mathcal{R}=S[\mathcal{R}_0+\mathcal{R}_1+\mathcal{R}_2+\mathcal{O}(S^{1/2})]\; ,
\end{align}
where $\mathcal{R}_0$ corresponds to a constant term and does not contribute to the scattering, $\mathcal{R}_1$ and $ \mathcal{R}_2$  describe, respectively, one-magnon and two magnon scatterings.
 The one-magnon  Raman scattering essentially probes only the magnon excitations in the center of the Brillouin zone ($\vec{q=0}$), so  in terms of the Bogoliubov quasiparticle operators ${\bf y}_\vec{q=0}$  the one-magnon Raman  operator can be  written as in \cite{Yang2021}: 
$
\mathcal{R}_1 \sim  \hat{\bf{M}}^{(1)}({\vec{q}}=0)  \cdot {\bf y}_\vec{q=0},
$
where $\hat{\bf{M}}^{(1)}({\vec{q}}=0) = \vec{V}^{(1)}\cdot \vec{T}_\vec{q=0}$, where $\hat{\vec{V}^{(1)}}$ is a $1\times(2\mc{N}_m)$ vector of combinations  of the products of the components of the polarization matrices and coupling constants. At zero temperature, only the terms involving $b^\dagger_\mu$ operators  contribute, so since only the processes with a magnon creation on the branch $\mu$ is allowed. Therefore, 
\be
\mc{I}_1(\Omega) \sim \sum_{\mu} |\hat{\bf{M}}^{(1)}({\vec{q}}=0)_{\mc{N}_m+\mu}|^2 ~\delta(\Omega-\omega_{\mu,\vec{q}=0}),
\ee
 where $\Omega=\omega_{\rm in}-\omega_{\rm out}$ is the total energy transferred to the system (in units of $\hbar=1$).  At finite temperature, the response will be modified by the proper inclusion of the Bose factors  $n_{\omega_{\mu,\vec{q}=0}}$ for creation and  $(1-n_{\omega_{\mu,\vec{q}=0}})$ for annihilation operators. 
 
  In Figure~3 from the main text, we show the low-energy one-magnon Raman response in  the $(\mathbf{c},\mathbf{a}-\mathbf{b})$ polarization channel   computed with the Raman coupling parameters  estimated  
   for the incoming photon frequency $\omega_{\text{in}}=2.33$ eV  corresponding to the $532$ nm laser used  here in the experimental Raman setup.  We can calculate explicitly $J^{(2)}=-316.324$ meV, $K^{(2)}=465.287$ meV, $K^{(4)}=-2865.05$ meV, $K'^{(4)}=-797.698$ meV, $J'^{(4)}=876.358$ meV, $\Gamma^{(3)}=-1333.2$ meV and $h_{\Gamma}^{(3)}=-436.864$ meV according to Ref.\cite{Yang2021}. 
   Then the one-magnon Raman intensity is computed by taking the Lorentizan form of $\delta(x)\rightarrow\frac{1}{\pi}\frac{\eta}{x^2+\eta^2}$ with the broadening factor $\eta=0.4$. 
   We first compute the Raman intensity  with the LSW spectrum obtained in 
   $\mathcal{H}_2$, and then repeat the same calculation with the  renormalized spin excitations obtained by diagonalizing  $\mathcal{H}'_2$.
    In both calculations, we obtain only one sharp one-magnon peak despite there are two low-energy $\mathbf{q}=0$ modes in the calculated energy spectrum. There are several possible origins of the missing second peak. One possibility is that 
    both calculations do not take properly into account the magnon-magnon interactions caused by the anharmonic terms. Another possibility  might come from the difference between  the actual  ground state of $\beta$-Li${}_2$IrO${}_3$ and the approximate commensurate magnetic state which we consider in our calculations. In  both cases, the approximation of non-interacting  magnon modes will not be sufficient.  
    
    At last, we use a {\it phenomenological} model, in which we allow for the minimal
    interaction between the low-energy  magnon modes, to nicely explain the experimental findings.
In particular, we  mimic the overall effect of the magnon-magnon interactions by adding some  off-diagonal elements to the diagonalized Hamiltonian of (\ref{H2Diag}). Since we focus on the low-energy  $\mathbf{q}=0$ excitations,   a minimally coupled Hamiltonian between the first and second excited modes can be obtained by using two real coupling constants $A$ and $B$ as
\begin{align}
    \mathcal{H}_{2,\mathbf{q}=0}^c={\bf y}_{\vec{q}=0}^\dagger \begin{pmatrix}
    \omega_{1,\mathbf{q}=0} & A &\cdots & B & B & \cdots\\
    A & \omega_{2,\mathbf{q}=0} &\cdots & B & B & \cdots\\
    \vdots & \vdots & \ddots & \vdots  & \vdots &\ddots\\
    B & B &\cdots& \omega_{1,\mathbf{q}=0} & A & \cdots\\
    B & B &\cdots & A & \omega_{2,\mathbf{q}=0} & \cdots\\
    \vdots & \vdots & \vdots & \vdots  & \vdots & \ddots
    \end{pmatrix}{\bf y}_{\vec{q}=0}.
\end{align}
The diagonalization of $\mathcal{H}_{2,\mathbf{q}=0}^c$ gives us another set of Bogoliubov quasiparticle operators $\mathbf{z}_{\mathbf{q}=0}=(c_{1,\mathbf{q}=0},\dots,c_{48,\mathbf{q}=0},c^\dagger_{1,-\mathbf{q}=0},\dots,c^\dagger_{48,-\mathbf{q}=0})^T$ with the relation $\mathbf{x}_{\mathbf{q}=0}=\mathbf{T}'_{\mathbf{q}=0}\cdot\mathbf{z}_{\mathbf{q}=0}$. Then we can recompute the one-magnon Raman response by $\hat{\bf{M}}^{(1)'}({\vec{q}}=0) = \vec{V}^{(1)}\cdot \vec{T}'_\vec{q=0}$. We can determine the numerical values of $A$ and $B$ by matching our numerical calculation to the experimental Raman intensity. In the inset of Figure~3 from the main text, we can see that by choosing $A=3.25$ meV and $B=1.9$ meV we achieve a very good agreement on the low-energy one-magnon Raman response, where two low-energy one-magnon  modes are visible in the  Raman response.

\end{document}